\documentclass[11pt]{article}
\usepackage{moriond,epsfig}

\def\Journal#1#2#3#4{{#1} {\bf #2}, #3 (#4)}

\def\PRL{\em Phys. Rev. Lett.}
\def\PRC{{\em Phys. Rev.} C}

\def\be{\begin{equation}}
\def\ee{\end{equation}}
\def\bea{\begin{eqnarray}}
\def\eea{\end{eqnarray}}

\begin{document}
\vspace*{4cm}
\title{IDENTIFIED PARTICLE PRODUCTION AT HIGH TRANSVERSE MOMENTUM IN NUCLEUS-NUCLEUS COLLISIONS AT RHIC}

\author{ V.S. Pantuev }

\address{Department of Physics and Astronomy, Stony Brook University,\\
Stony Brook, New York, 11794-3800, USA}

\maketitle\abstracts{Experimental results for identified hadron spectra 
in AA collisions at RHIC are reviewed. Jet suppression in 
central AuAu collisions at 200 GeV is seen in leading meson and baryon spectra
 at 
high $p_t$. Enhanced 
baryon production at intermediate $p_T$ ($2<p_t<5$ GeV/c) is related to quark 
content but not its mass. In d+Au collisions Cronin enhancement is observed 
with larger magnitude for baryons than for mesons, but this difference can 
not explain the
baryon-to-meson ratio in Au+Au collisions. In Au+Au  the observations at 62.4 GeV 
and 200 GeV are very similar, but  smaller suppresion of the leading particles 
is seen at low beam energy, 
which is evidence for a smooth beam energy dependence of the effect. Near and 
away side jets seen in two-particle correlations are very similar for the 
leading 
baryon and mesons.
}

\begin{figure}[ht]
\begin{minipage}[t]{8cm}
\psfig{figure=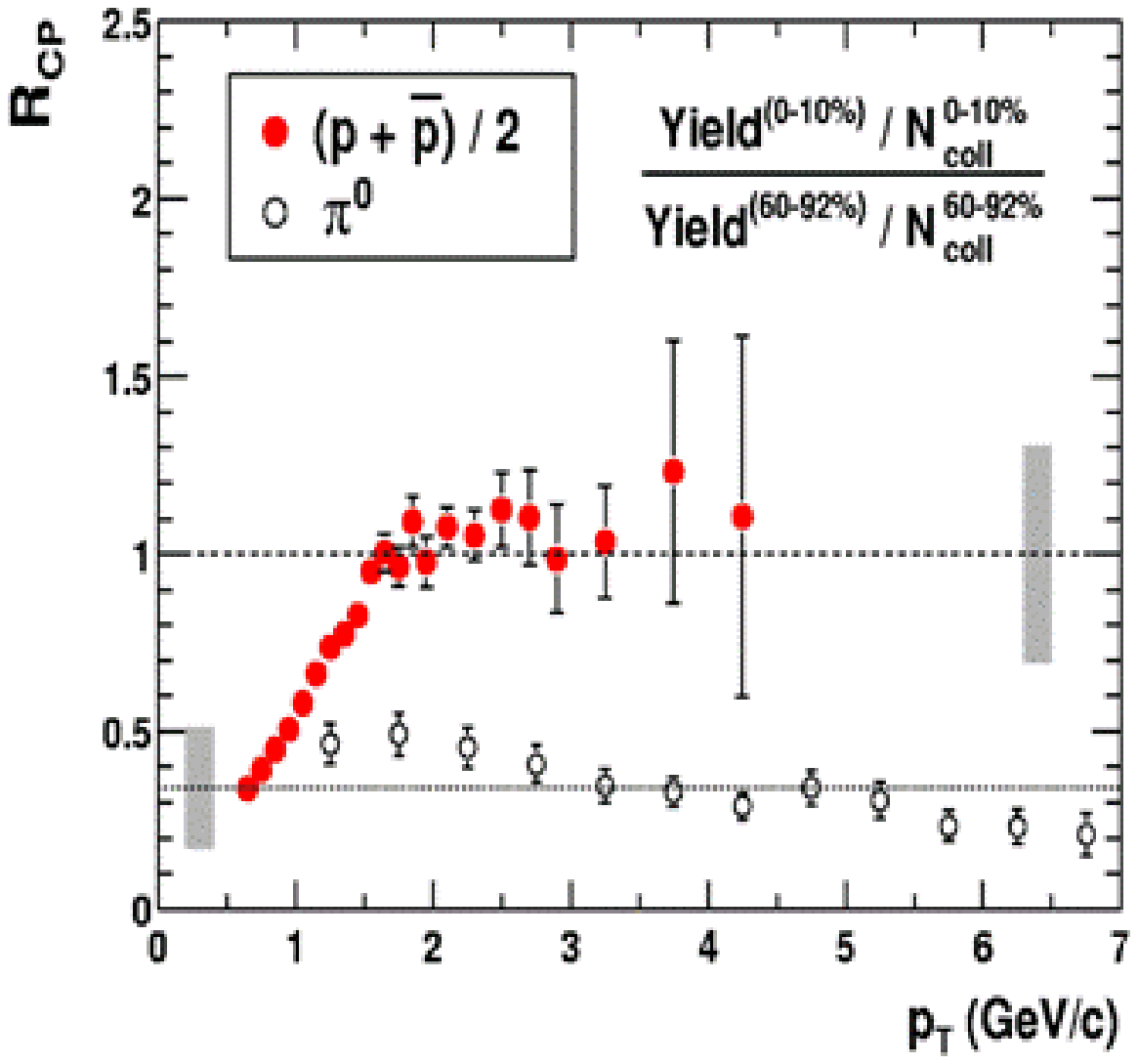,height=2.35in,width=7cm}
\caption{Central to peripheral yield ratio per  binary collision, $R_{cp}$, vs. 
transverse momentum for neutral pions and protons in Au+Au collisions at 200 
GeV. }
\label{fig:fig1}
\end{minipage}
\hfill
\begin{minipage}[t]{7cm}
\psfig{figure=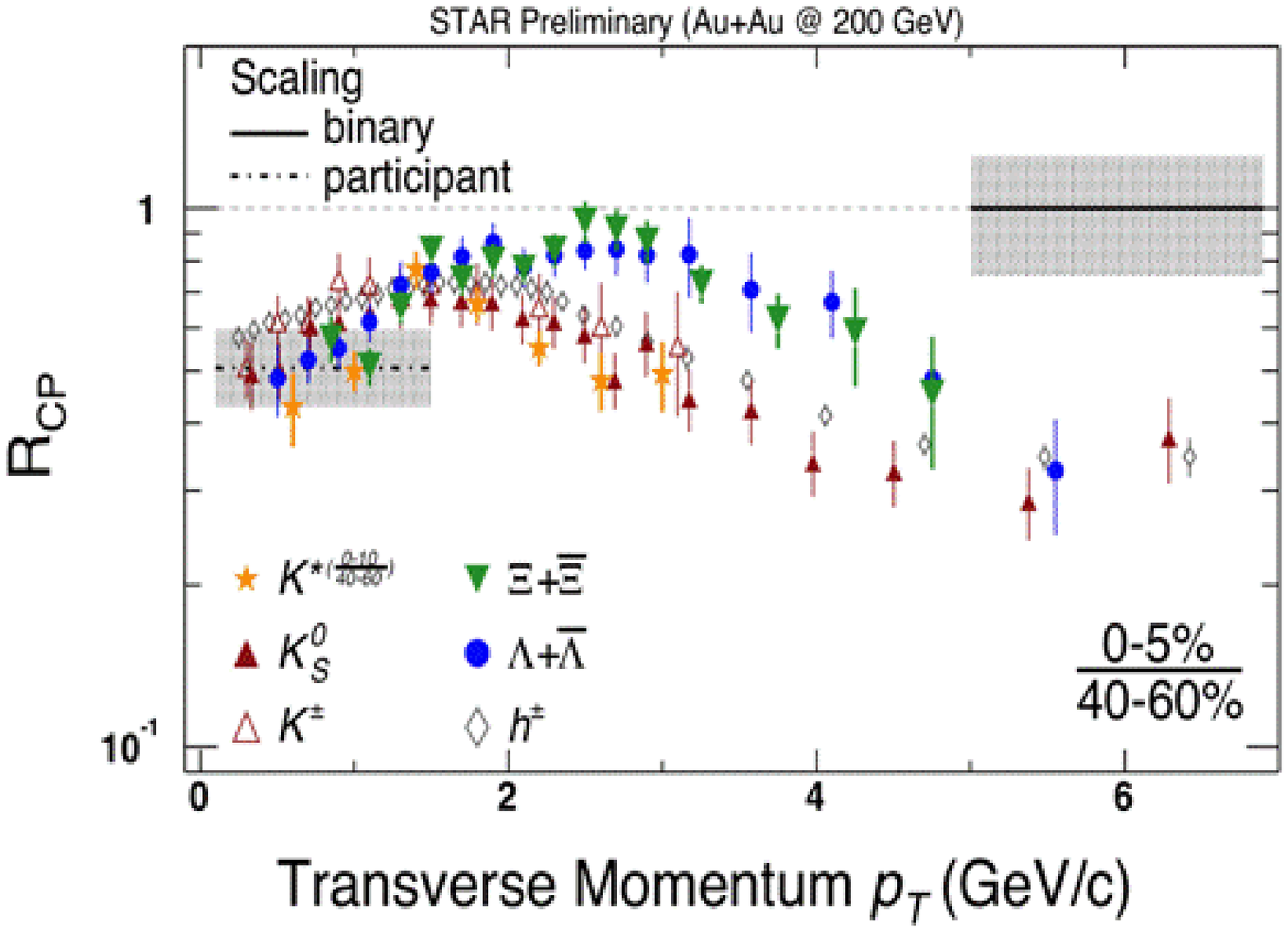,height=2.5in,width=7cm}
\caption{Central to peripheral yield ratio per collision vs. transverse momentum for charge mesons and baryons in Au+Au collisions at 200 GeV. }
\label{fig:fig3}
\end{minipage}
\end{figure}

Since the year 2000 four experiments at the Relativistic Heavy Ion Collider (RHIC) at 
Brookhaven National Laboratory are searching for a new state of nuclear matter: the quark-gluon plasma. All experiments confirmed the effect of particle suppression at
 high transverse momenta in the most central Au+Au collisions~\cite{hemmick}. This 
phenomenon is consistent with the general expectation from pQCD calculations for final-state partonic energy loss or ``jet quenching'' in produced dense matter. At the 
same time it was found that protons and antiprotons are not suppressed as 
pions~\cite{pscaling}, 
Figure~\ref{fig:fig1}. Here the parameter $R_{cp}$ is defind as:

\begin{equation}
R_{cp}(p_T) 
= \frac{(1/\langle N_{binary}^{cent} \rangle) \; d^{2}N^{A+A}_{cent}/dp_T d\eta }
{(1/\langle N_{binary}^{periph} \rangle)\; d^{2}N^{A+A}_{periph}/dp_T d\eta }.
\label{eq:Rcp_defined}
\end{equation}

\begin{figure}[ht]
\begin{minipage}[t]{8cm}
\psfig{figure=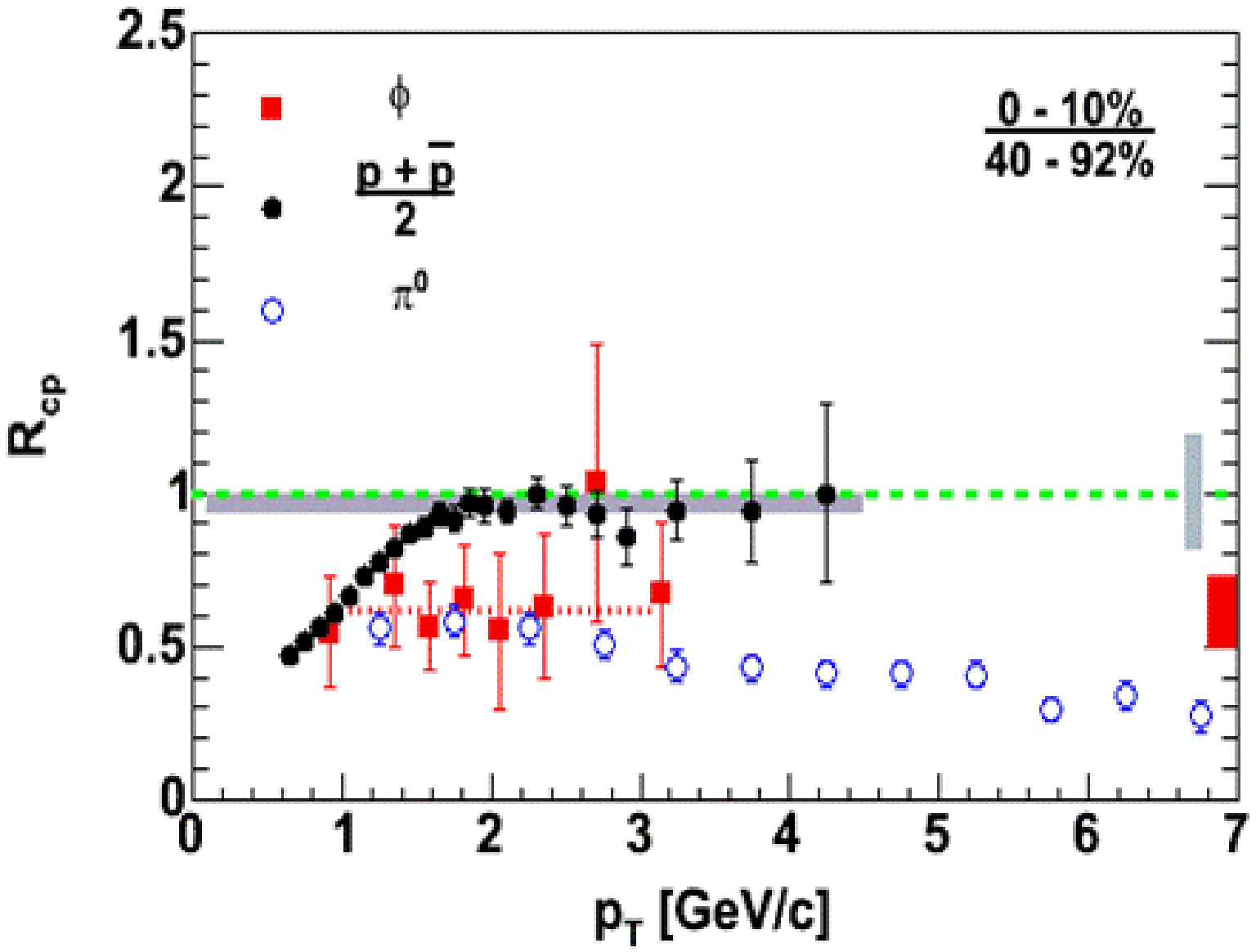,height=2.5in}
\caption{Central to peripheral yield ratio per collision vs. transverse momentum for 
pions, protons and $\phi$-mesons in Au+Au collisions at 200 GeV.}
\label{fig:fig4}
\end{minipage}
\hfill
\begin{minipage}[t]{7.5cm}
\psfig{figure=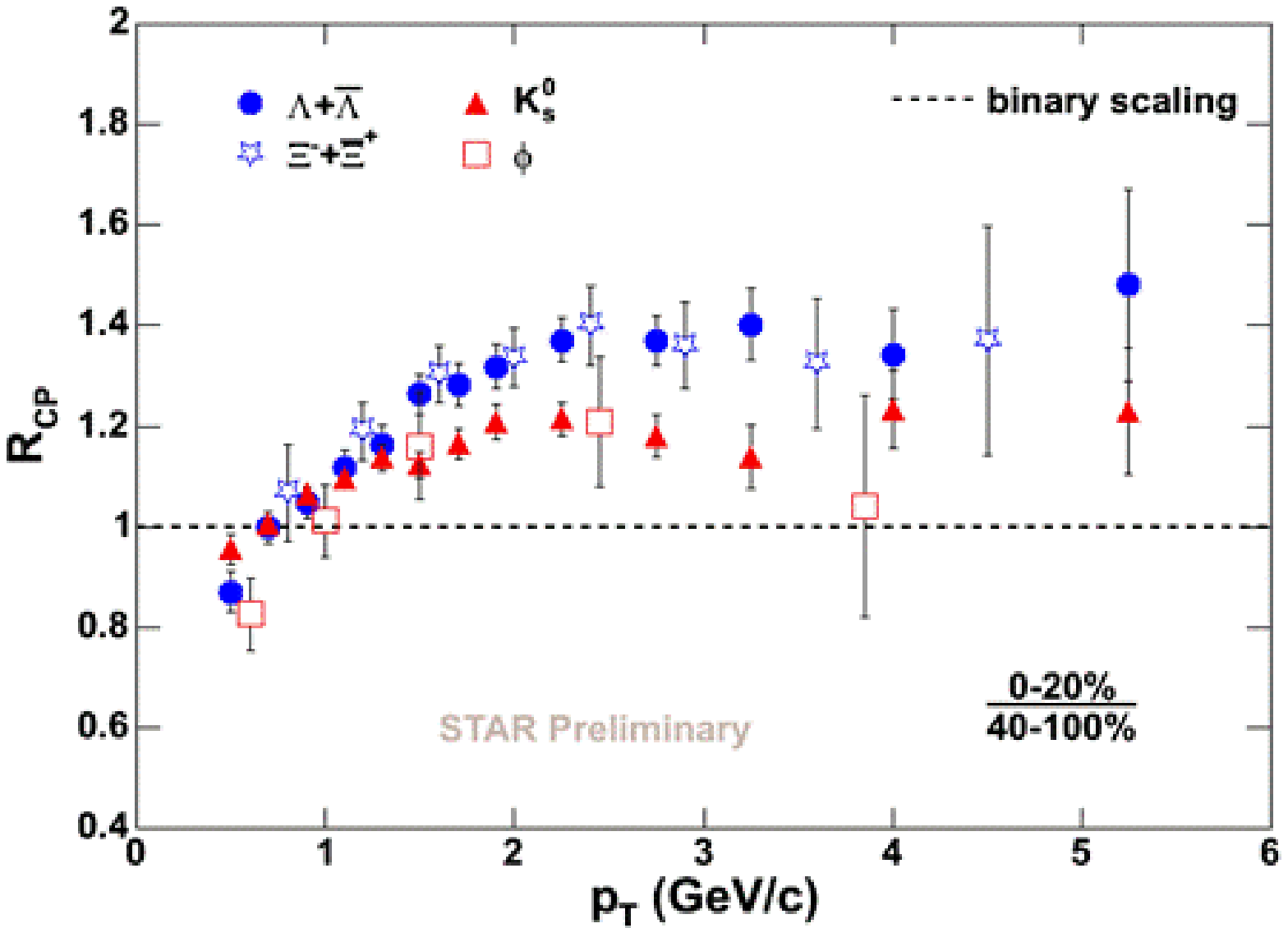,height=2.3in, width=7cm}
\caption{Central to peripheral yield ratio per  binary collision vs. transverse momentum for hyperons, kaons and $\phi$-mesons in d+Au collisions at 200 GeV. Dotted line at $R_{CP}=1$ shows expectation from binary scaling.}
\label{fig:fig6}
\end{minipage}
\end{figure}

Jet production, and hence its leading particle, has a small cross 
section and should be proportional to the number of binary nucleon-nucleon 
collisions, $N_{binary}$, for each AA collision centrality. Thus, to see the nuclear 
modification effect versus centrality, particle yields in $R_{cp}$ are normalized 
to $N_{binary}$.\\
To understand whether the proton non-suppression is a  mass effect due to an additional radial 
flow or a pure quark-content effect (byryon vs. meson), many measurents have been done with other 
particles~\cite{starhyperons,phi}, Figure~\ref{fig:fig3} and Figure~\ref{fig:fig4}. First, data confirm that this is a baryon effect, probably related to the number of 
constituent quarks, three for baryons and two for mesons. Second, baryon suppression is also observed, but it starts  at higher momenta above 4 GeV/c. The baryon enhancement seen in Au+Au at 2-4 GeV/c is a feature of the   
transision region between soft and hard scattering processes.

\begin{figure}[htb]
\begin{center}
\begin{minipage}[t]{7.5cm}
\psfig{figure=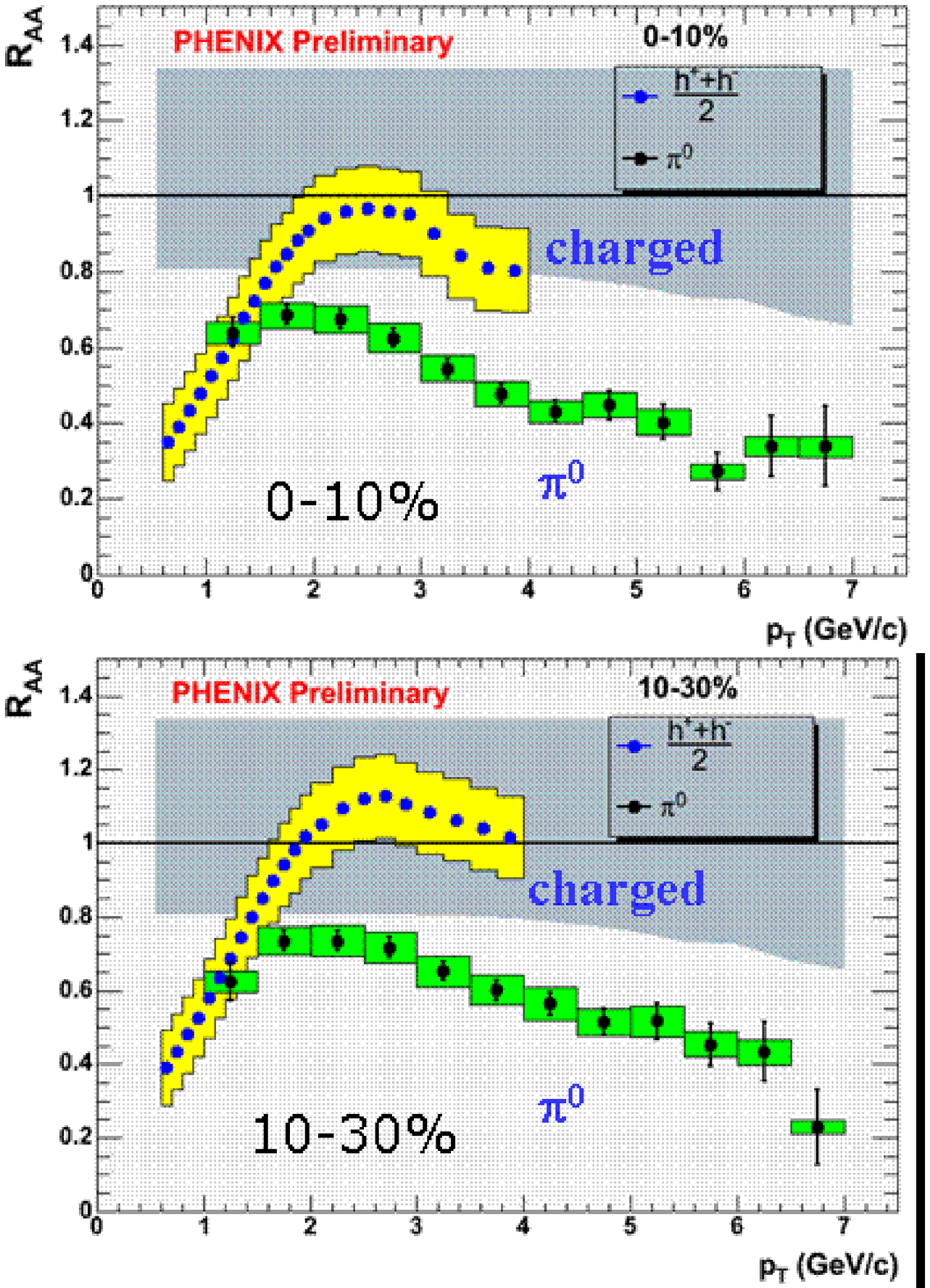,height=2.3in,width=7cm}
\caption{Nuclear modification factor at 62.4 GeV for most central and 
semi-central  Au+Au collisions. Top points - for all charged hadrons, 
lower points - for neutral pions only.}
\label{fig:fig7}
\end{minipage}\hfill
\begin{minipage}[t]{8cm}
\psfig{figure=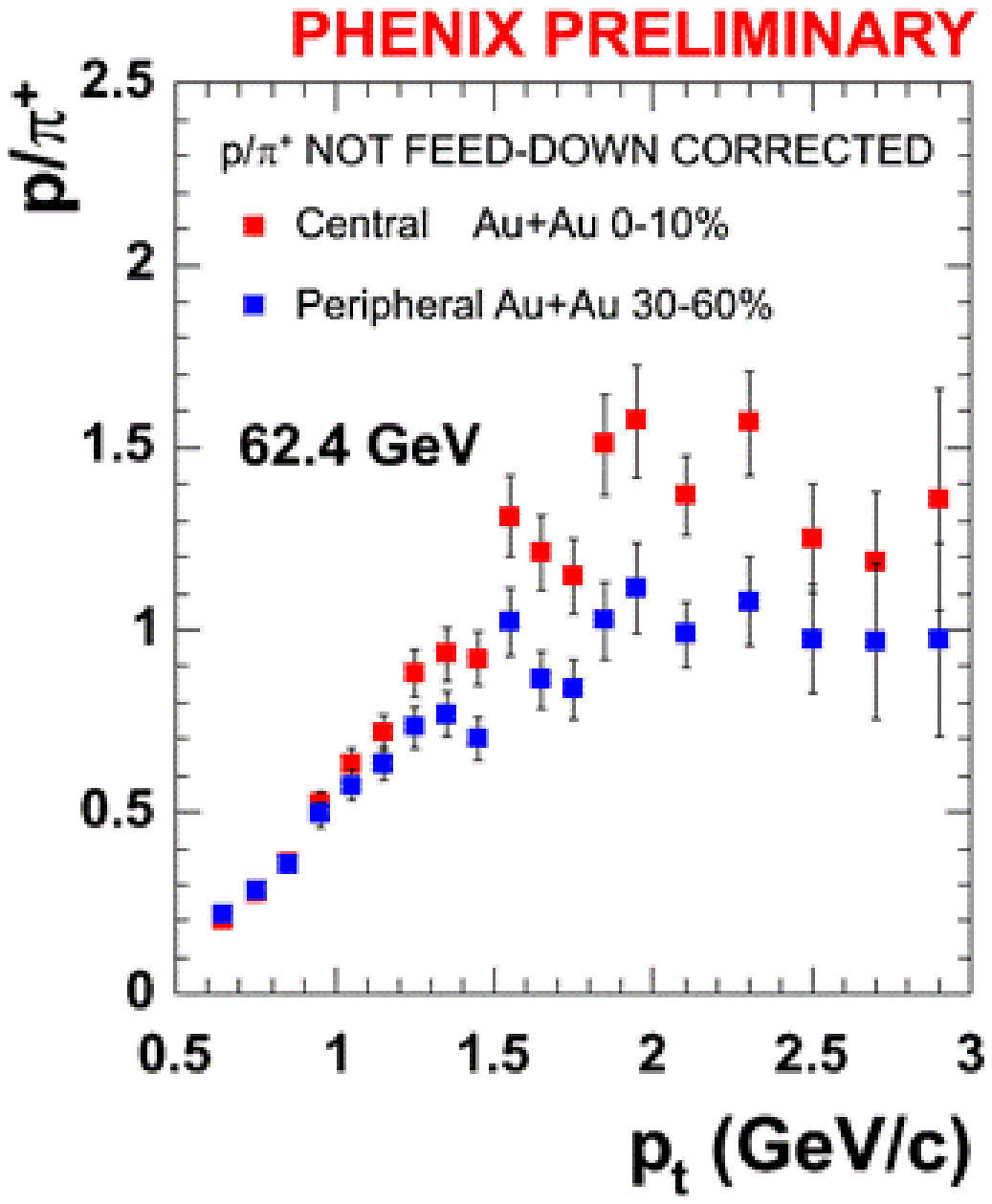,height=2.5in,width=8cm}
\caption{Proton to pion ratio at 62.4 GeV for most central and semi-central  Au+Au collisions.}
\label{fig:fig8}
\end{minipage}
\end{center}
\end{figure}

To confirm that jet quenching is indeed a final state effect in central 
Au+Au collisions, a set of ``controll'' experiments at RHIC has been performed with d+Au 
beams. If the suppression is an initial state effect,
then it should be seen in low density d+Au collisions also. At midrapidity no suppression was observed. 
Even more, some enhancement was seen for more central d+Au 
events~\cite{stardAu}, 
Figure~\ref{fig:fig6}. This, so called, Cronin effect was observed already in 
mid 1970-th in proton beams on nuclear targets. The enhancement is larger for baryons than 
for mesons but this is a small effect and can not explain the factor 4-5 
baryon-meson difference at intermediate $p_t$ in  central Au+Au collisions.\\
The observation of strong meson suppression at high transverse momenta in 
Au+Au collisions at 200 GeV gave rise to the question about the beam energy dependance of 
a such effect. RHIC performed a short one week Au+Au run at 
$\sqrt{s}=62.4$ GeV. 
Preliminary results, Figure~\ref{fig:fig7}, show a significant suppression of 
the neutral pion yield for the most central collisions~\cite{bathe62}, but the 
suppression factor at 62.4 GeV is less than at 200 GeV. This is evidence 
for a smooth beam energy dependence of the effect. At the same time, 
at 62.4 GeV the nuclear modification factor for charged hadrons, i.e. pions, 
kaons and protons, does not show suppression in the measured range 
of  $p_t<4$ GeV/c. An explanation of this feature is presented in
 Figure~\ref{fig:fig8}: in most central events at high momentum the charged 
hadron yield  contains more protons than pions. There is no direct measurements
of the nuclear modification factor for identified baryons at $\sqrt{s}=62.4$ GeV, 
but from  Figure~\ref{fig:fig7} and Figure~\ref{fig:fig8} one may conclude, 
that baryon production, or at least for protons, is similar at 
62.4 and 200 GeV.\\

\begin{figure}[htb]
\begin{minipage}[t]{7.5cm}
\psfig{figure=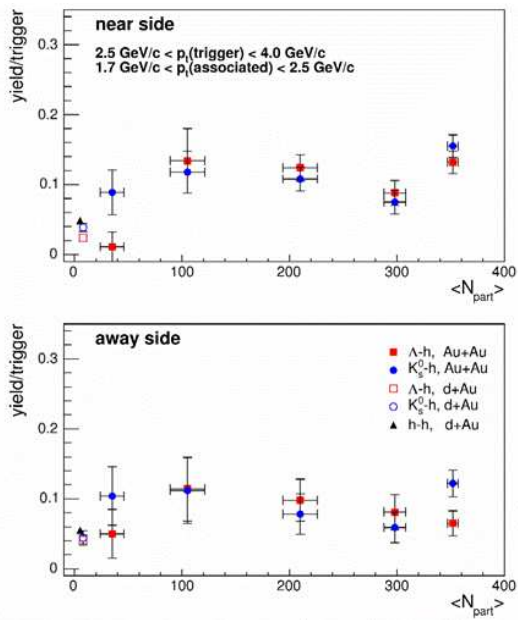,width=7.3cm}
\end{minipage}
\hfill
\begin{minipage}[t]{8cm}
\psfig{figure=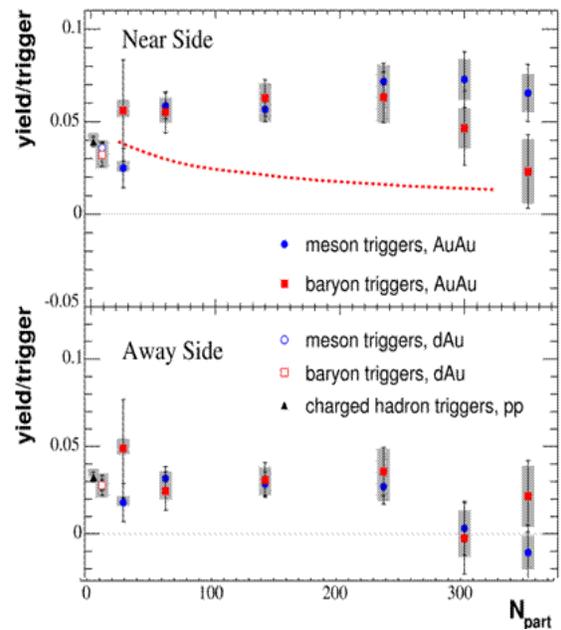,height=8.6cm,width=8cm}
\end{minipage}
\caption{Results from 2-particle correlation with identified leading hadron: yield  for near and away side jets per trigger particle vs. number of participants. STAR preliminary results are on the left histograms, PHENIX results - on the right. In PHENIX, baryons were protons and antiprotons, mesons - pions and Kaons; dotted line shows the results of recombination model. Diffrence in STAR/PHENIX yields is due to about factor 2 larger preseudo-rapidity acceptance in STAR.}
\label{fig:fig10_fig11}
\end{figure}

The very large particle multiplicity in Au+Au collisions at RHIC does not allow for a full jet reconstruction - the background of soft particles is too high. 
Nevertheless, it was demonstrated that two particle azimuthal correlation for 
large transverse momentum particles could be a valid tool for jet 
identification~\cite{starjet}. 2-particle azimuthal correlations allow to 
investigate the centrality dependence, 
the shape and yield in forward and backward jets, to measure the spectrum of 
particles in the jet. The results for the particle yield 
in the  jets~\cite{starcorr,phenixcorr} are presented in 
Figure~\ref{fig:fig10_fig11}. The leading particle was identified as a meson or 
baryon, the  
second particle was a non-identified charged hadron. The particle yield per trigger in the near side and away side jets is  
almost the same for identified leading mesons and baryons, at least in the 
mesured momentum range, regardless of a large difference in the inclusive 
 baryon and meson yields  at  $2<p_t<5$ GeV/c. \\ 
Different versions of the recombination/coalescence  parton model were 
proposed to 
describe the enhanced $p/\pi$ ratio for inclusive spectra~\cite{fries}. 
This model goes beyond the standard picture of  high 
transverse momentum hadron production by fragmentation of the energetic parton. It proposes that 
hadron production at a few GeV/c in an enviroment with  high density of partons 
occurs via the parton recombination. Another semi-phenomenological model describes 
the large hadron/pion ratio by a non-universal parton transverse momentum 
broadening~\cite{zhang}. Authors of this paper argue that smearing of the 
parton intrinsic $k_T$ in pA and AA collisions should be larger for protons
than for pions. Less schematic calculations should be done to test   
both models.\\
In conclusion, we have to state that the nature of big difference in baryon and meson production in Au+Au collisions at RHIC in transision region between soft and hard scattering is still not clear, more data are needed, more consistent 
explanations are required. 

\section*{References}
\begin{thebibliography}{99}
\bibitem{hemmick}T.K. Hemmick, J. Phys. {\bf G30}, S659 (2004).

\bibitem{pscaling}S.S Adler {\it et al} (PHENIX Collaboration, \Journal{\PRL}{91}{172301}{2003}.

\bibitem{starhyperons}K. Scheweda (STAR Collaboration), J. Phys. {\bf G30}, S693 (2004).
\bibitem{phi}S.S Adler {\it et al} (PHENIX Collaboration), nucl-ex/0410012, (2004).

\bibitem{stardAu}L.S Barnby (STAR Collaboration), J. Phys. {\bf G30}, S1121 (2004).

\bibitem{bathe62}S. Bathe (PHENIX Collaboration), talk at APS Metting, October 27-30, Chicago, IL, USA.


\bibitem{starjet}C. Adler {\it et al} (STAR Collaboration), \Journal{\PRL}{90}{082302}{2003}.

\bibitem{starcorr}J. Belchikova (STAR Collaboration),  talk at APS Metting, October 27-30, Chicago, IL, USA. 

\bibitem{phenixcorr}S.S Adler {\it et al} (PHENIX Collaboration), nucl-ex/0408007, (2004).

\bibitem{fries}R.J. Fries  {\it et al},  \Journal{\PRL}{90}{202303}{2003}. R.C. Hwa and C.B. Yang,  \Journal{\PRC}{70}{024904}{2004}.


\bibitem{zhang}X. Zhang and G. Fai, hep-ph/0306227 (2004).

\end{thebibliography}
\end{document}